\documentclass[12pt,twoside,fleqn]{article}
\usepackage{epsfig}
\def\lsim{\raise0.3ex\hbox{$<$\kern-0.75em\raise-1.1ex\hbox{$\sim$}}}
\def\gsim{\raise0.3ex\hbox{$>$\kern-0.75em\raise-1.1ex\hbox{$\sim$}}}
\makeatletter
\@addtoreset{equation}{section}
\makeatother

\newcommand{\beqn}{\begin{equation}}
\newcommand{\eqn}{\end{equation}}
\newcommand{\bqa}{\begin{eqnarray}}
\newcommand{\eqa}{\end{eqnarray}}
\newcommand{\bqas}{\begin{eqnarray*}}
\newcommand{\eqas}{\end{eqnarray*}}
\newcommand{\bdm}{\begin{displaymath}}
\newcommand{\edm}{\end{displaymath}}

\newcommand{\tr}{\mbox{Tr~}}
\newcommand{\re}{\mbox{Re~}}
\newcommand{\nn}{\nonumber}
\newcommand{\plaq}{\mbox{\raisebox{-1.25mm}
{\epsfig{file=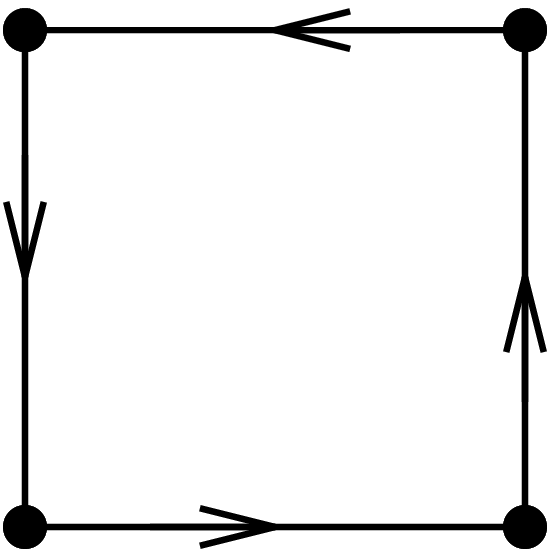,height=5mm
}}~}}
\newcommand{\loOp}{\mbox{\raisebox{-1.25mm}
{\epsfig{file=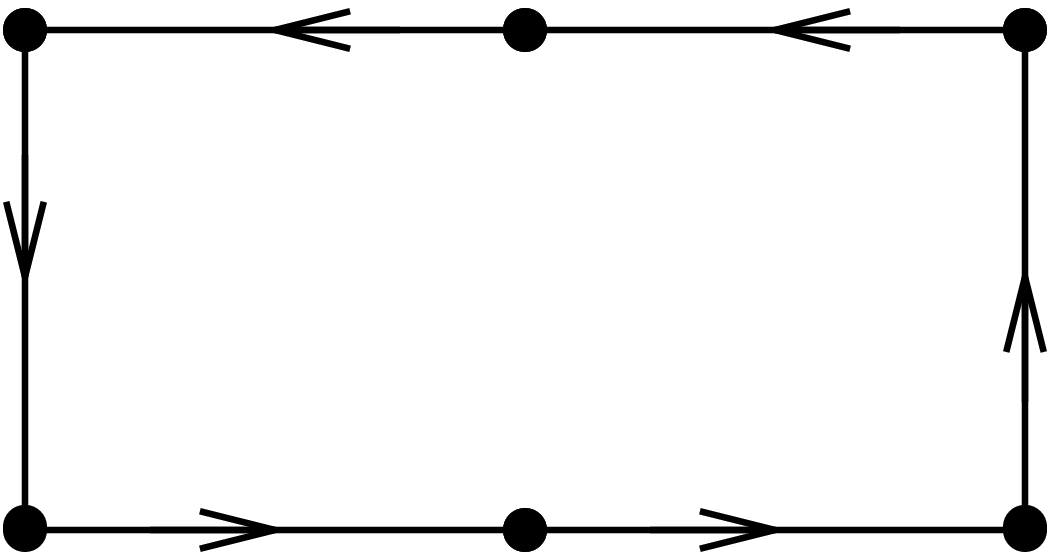,height=5mm
}}~}}
\newcommand{\lOop}{\mbox{\raisebox{-4mm}
{\epsfig{file=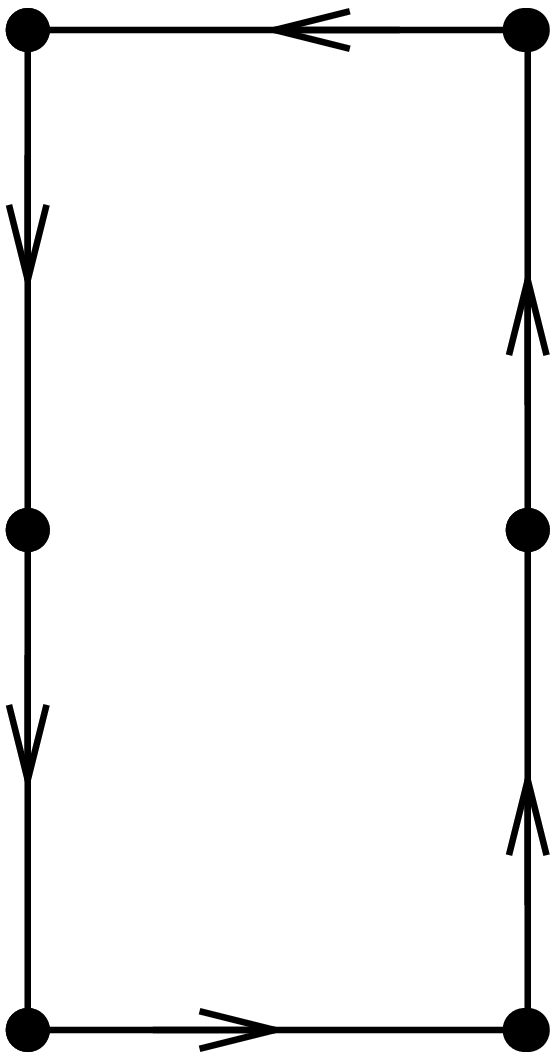,height=10mm
}}~}}
\newcommand{\alink}{\mbox{
\begin{picture}(2.5,.2)
\linethickness{1mm}
\multiput(0,0.1)(2,0){2}{\circle*{0.1}}
\multiput(1.01,0.1)(0,0){1}{\circle{0.2}}
\put(0.8,0.){\link}
\put(-0.2,0.){\linka}
\put(1.1,-.2){\scriptsize \( x \) }
\put(0,-.2){\scriptsize \( y \) }
\put(2.0,-.2){\scriptsize \( y \) }
\end{picture}}}
\newcommand{\alinkfat}{\mbox{
\begin{picture}(3.0,.2)
\linethickness{1mm}
\multiput(0,0.1)(2.4,0){2}{\circle*{0.1}}
\multiput(1.21,0.1)(0,0){1}{\circle{0.2}}
\put(1.2,-1.){\staple}
\put(-0.2,-1.){\staplea}
\put(1.1,-.3){\scriptsize \( x \) }
\put(-.4,-.2){\scriptsize \( y \) }
\put(2.6,-.2){\scriptsize \( y \) }
\end{picture}}}
\newcommand{\blinkba}{\mbox{
\begin{picture}(2.5,2.5)
\thicklines
\multiput(1,-1)(0,-1){2}{\circle*{0.1}}
\multiput(1,0)(0,0){1}{\circle{0.2}}
\multiput(2,0.0)(0,1){3}{\circle*{0.1}}
\multiput(0,-2)(0,0){1}{\circle*{0.1}}
\put(1,-2.0){\vector(-1,0){1}}
\put(1,-1.0){\vector(0,-1){1}}
\put(1,0.0){\vector(0,-1){1}}
\put(1,0.0){\vector(1,0){1}}
\put(2,0.0){\vector(0,1){1}}
\put(2,1.0){\vector(0,1){1}}
\put(0.5,-0.1){\scriptsize \( x \) }
\put(1.5,2.0){\scriptsize \( y \) }
\put(-0.4,-2.1){\scriptsize \( y \) }
\end{picture}}}
\newcommand{\blinkbb}{\mbox{
\begin{picture}(2.5,2.5)
\thicklines
\multiput(0,0.0)(0,-1){3}{\circle*{0.1}}
\multiput(1,1.0)(0,1){2}{\circle*{0.1}}
\multiput(1,0.0)(0,){1}{\circle{0.2}}
\multiput(2,2)(0,1){1}{\circle*{0.1}}
\put(0,-1.0){\vector(0,-1){1}}
\put(0,0){\vector(0,-1){1}}
\put(1,0){\vector(-1,0){1}}
\put(1,0){\vector(0,1){1}}
\put(1,1){\vector(0,1){1}}
\put(1,2){\vector(1,0){1}}
\put(1.25,-0.1){\scriptsize \( x \) }
\put(2.2,2.0){\scriptsize \( y \) }
\put(-0.4,-2.1){\scriptsize \( y \) }
\end{picture}}}
\newcommand{\blinkbc}{\mbox{
\begin{picture}(2.5,2.5)
\thicklines
\multiput(1,1)(0,1){2}{\circle*{0.1}}
\multiput(1,0)(0,0){1}{\circle{0.2}}
\multiput(2,0)(0,-1){3}{\circle*{0.1}}
\multiput(0,2)(0,1){1}{\circle*{0.1}}
\put(1,2){\vector(-1,0){1}}
\put(1,1){\vector(0,1){1}}
\put(1,0){\vector(0,1){1}}
\put(1,0){\vector(1,0){1}}
\put(2,0){\vector(0,-1){1}}
\put(2,-1){\vector(0,-1){1}}
\put(0.5,-0.1){\scriptsize \( x \) }
\put(1.5,-2.1){\scriptsize \( y \) }
\put(-0.5,2.0){\scriptsize \( y \) }
\end{picture}}}
\newcommand{\blinkbd}{\mbox{
\begin{picture}(2.5,2.5)
\thicklines
\multiput(0,0)(0,1){3}{\circle*{0.1}}
\multiput(1,-1)(0,-1){2}{\circle*{0.1}}
\multiput(1,0)(0,-1){1}{\circle{0.2}}
\multiput(2,-2)(0,1){1}{\circle*{0.1}}
\put(0,1){\vector(0,1){1}}
\put(0,0){\vector(0,1){1}}
\put(1,0){\vector(-1,0){1}}
\put(1,0){\vector(0,-1){1}}
\put(1,-1){\vector(0,-1){1}}
\put(1,-2){\vector(1,0){1}}
\put(1.25,-0.1){\scriptsize \( x \) }
\put(2.1,-2.1){\scriptsize \( y \) }
\put(-0.5,2.0){\scriptsize \( y \) }
\end{picture}}}
\newcommand{\staple}{\mbox{
\begin{picture}(1.2, 2.2)
\thicklines
\put(0,1.2){\vector(0,1){1}}
\put(0,1){\vector(0,-1){1}}
\put(0,2.2){\vector(1,0){1}}
\put(0,0.){\vector(1,0){1}}
\put(1,0.){\vector(0,1){1}}
\put(1,2.2){\vector(0,-1){1}}
\end{picture}}}
\newcommand{\staplea}{\mbox{
\begin{picture}(1.2, 2.2)
\thicklines
\put(1,1.2){\vector(0,1){1}}
\put(0,2.2){\vector(0,-1){1}}
\put(1,2.2){\vector(-1,0){1}}
\put(1,1){\vector(0,-1){1}}
\put(1,0){\vector(-1,0){1}}
\put(0,0){\vector(0,1){1}}
\end{picture}}}
\newcommand{\link}{\mbox{
\begin{picture}(1.1, .1)
\thicklines
\put(0,0.1){\vector(1,0){1}}
\end{picture}}}
\newcommand{\linka}{\mbox{
\begin{picture}(1.1, .1)
\thicklines
\put(1,0.1){\vector(-1,0){1}}
\end{picture}}}

\begin{document}
\thispagestyle{empty}
%
 \mbox{} \hfill CERN-TH/2000-037\\
 \mbox{} \hfill BI-TP 2000/01\\
\begin{center}
{{\large \bf The Pressure in 2, 2+1 and 3 Flavour QCD}
 } \\
\vspace*{1.0cm}
{\large F. Karsch$^{1,2}$, E. Laermann$^1$ and A. Peikert$^{1,3}$}

\vspace*{1.0cm}
{\normalsize
$\mbox{}$ {$^1$Fakult\"at f\"ur Physik, Universit\"at Bielefeld,
D-33615 Bielefeld, Germany}\\
$\mbox{}$ {$^2$Theory Division, CERN, CH-1211, Geneva, Switzerland} \\
$\mbox{}$ {$^3$ Department of Physics, Hiroshima University,
Higashi-Hiroshima 739-8521, Japan}
}
\end{center}
\vspace*{1.0cm}
\centerline{\large ABSTRACT}

\baselineskip 20pt

\noindent
We calculate the pressure in QCD with two and three light quarks
on a lattice of size $16^3\times 4$ using tree level improved gauge
and fermion actions. We argue that for temperatures $T\; \gsim \; 2T_c$
systematic effects due to the finite lattice cut-off and non-vanishing
quark masses are below 15\% in this calculation and give an estimate
for the continuum extrapolated pressure in QCD with massless
quarks. We find that the flavour dependence of the pressure
is dominated by that of the Stefan-Boltzmann constant.
Furthermore we perform a calculation of the pressure using 2
light ($m_{u,d}/T=0.4$) and one heavier quark ($m_s/T = 1$). In this case
the pressure is reduced relative to that of three flavour QCD. This
effect is stronger than expected
from the mass dependence of an ideal Fermi gas.
\vfill
\eject
\baselineskip 15pt

\section{Introduction}

One of the central goals in studies of the QCD thermodynamics on the
lattice is the calculation of the equation of state for QCD with
a realistic mass spectrum. Controlling the influence of a heavier strange
quark on the QCD phase transition
as well as understanding its contribution to bulk thermodynamic observables,
e.g. the pressure or energy density, is of fundamental importance
for the analysis and interpretation of heavy ion experiments which look for
signals from the quark-gluon plasma phase of QCD.
While the former problem is closely related to the chiral structure of
QCD and requires numerical computations with light quarks close to the chiral
limit \cite{strangeTc} the latter question can be addressed in
computations with moderate quark masses already.

An increase in the relative abundance of strange particles at high temperature
and density is being discussed as a signature for the formation of a
quark-gluon plasma \cite{Raf99}. A first analysis of the contribution of the
strange quark sector to the energy density has been performed already some
time ago in a lattice calculation \cite{Kog88}.
This calculation suggested that even at temperatures a few times the critical
temperature the strange quark contribution to the overall energy density
is strongly suppressed. The energy density in the strange quark sector
was found to reach only half the value of a non-interacting Fermi gas.
This conclusion, however, has been drawn by separately analyzing operators
which in a non-interacting gas would describe the contribution of light and
heavy quarks, respectively, to the overall energy density. Such a separation
of different contributions is questionable at temperatures a few times $T_c$.
In fact, the experience gained from lattice calculations of
the equation of state in the pure gauge sector \cite{Boy96}, the failure
of a perturbative description of its high temperature behaviour \cite{Bra96}
as well as the success of hard thermal loop resummed perturbative
calculations \cite{And99a,Bla99} suggest that the high temperature phase of
QCD remains non-perturbative even at temperatures several times $T_c$.
This prohibits an isolated analysis of the contribution of
different parton sectors to bulk thermodynamic observables, e.g. the
free energy density $f(T)$ which in the thermodynamic limit yields the
pressure, $p(T) = -f(T)$. One rather has to analyze the variation of $f(T)$
with the number of flavours and its quark mass dependence to deduce the
effect of a non-vanishing strange quark mass on the thermodynamics.

The pioneering calculation of the strange quark contribution to the
QCD equation of state \cite{Kog88} uses the standard Wilson gauge and
staggered fermion actions. It still is strongly influenced by
cut-off effects and moreover, makes use of partly perturbative relations
in the calculation of thermodynamic quantities. Both problems can be
handled now much better. The use of the {\it integral method} \cite{Eng90}
allows an entirely non-perturbative calculation of thermodynamic observables.
Cut-off effects can be strongly reduced in calculations with improved
gauge \cite{Bei96,Pap96,Bei99,Oka99} and fermion \cite{Jos97,Hel99}
actions.

We will present here results from a calculation with improved gauge and
improved staggered fermion actions for two and three quarks of equal mass 
as well as two light and a heavier strange quark. Based on an analysis of 
the remaining cut-off effects and the quark mass dependence we will give an 
estimate for the free energy density ($\sim$ pressure) of QCD with massless 
quarks at temperatures $T\; \gsim \; 2 T_c$. In addition
we discuss the contribution of the heavier strange quark to the free
energy density in this temperature interval.

This paper is organized as follows.
In the next Section we will describe the lattice action used for our
calculations and discuss the cut-off dependence of thermodynamic
observables in the infinite temperature, ideal gas limit. In Section 3
we present details of our numerical calculation and give the basic
numerical data entering our analysis of the pressure which
is performed in Section 4. In Section 5 we discuss the extrapolation of
our results to the continuum limit for which we give a first estimate.
Our conclusions are given in Section 6.

\section{Improved action, Cut-off dependence}

The equation of state for QCD with two light quarks has been analyzed
recently on lattices with temporal extent $N_\tau=4$ \cite{Blu95}
and 6 \cite{Ber97}. In these calculations, which have been performed
with the standard Wilson gauge and staggered fermion actions, a
sizeable cut-off dependence has been observed. The general pattern,
a strong over-shooting of the continuum ideal gas limit, is in accordance
with the known cut-off effects for an ideal quark-gluon gas
calculated with these actions on lattices with finite temporal extent.
Also the first thermodynamic studies performed for four flavour QCD with
an improved fermion action \cite{Jos97} indicate that the qualitative
features of the cut-off dependence closely follow the pattern seen for
an ideal gas. Analyzing cut-off effects in the ideal gas limit
thus provides useful guidance for selecting an improved action for
thermodynamic calculations.

In our calculations we use a tree level, Symanzik improved gauge action,
which in addition to the standard Wilson plaquette term also includes the
planar 6-link Wilson loop, and a staggered fermion action with 1-link and
bended 3-link terms,
\begin{eqnarray}
Z(T,V) &=& \int \prod_{x,\mu}{\rm d} U_{x,\mu} {\rm e}^{-\beta S_G }
\prod_{f}\biggl(\int \prod_x {\rm d}\bar{\chi}_x
{\rm d}\chi_x\; {\rm e}^{-S_F(m_{f,L})} \biggr)^{1/4} \\
~~~~\nonumber \\
S_G &=& c_4\; S_{plaquette} + c_6\; S_{planar} \nonumber \\
&\equiv& \sum_{x, \nu > \mu}~ {5\over 3}~\left(
1-\frac{1}{3}\re\tr\plaq_{\mu\nu}(x)\right) \nn\\
& & -{1\over 6 }\left(1-\frac{1}{6}\re\tr
\left(\loOp_{\mu\nu}(x)+\lOop_{\mu\nu}(x)\right)\right)
\end{eqnarray}
\begin{eqnarray}
\lefteqn{ S_F (m_{f,L}) ~=~ c_1^F S_{1-link,fat} (\omega) +
c_3^F S_{3-link}+  m_{f,L}\sum_x
\bar{\chi}_x^f \chi_x^f \nn ~~~~~~~~~~~~~~~~~~~~~~~}\\
&\equiv& \sum_x \bar{\chi}_x^f~\sum_\mu ~ \eta_\mu(x) ~ \Bigg(
{3\over 8}~\Bigg[ \alink~ +~ \omega~~\sum_{\nu \ne \mu}~~ \alinkfat\Bigg]
\nn \\[4mm]
& & + {1\over96}~\sum_{\nu\ne \mu} ~\Bigg[ \blinkbd + \blinkbc ~+
 \blinkba + \blinkbb \Bigg] \Bigg) \chi_y^f \nn \\[13mm]
& & + m_{f,L}  \sum_x~\bar{\chi}_x^f \chi_x^f  \quad .
\end{eqnarray}
Here we have made explicit the dependence of the action on different
quark flavours, $f$, and the corresponding bare quark masses $m_{f,L}$
and give an intuitive graphical representation of the action.
With $\eta_\mu(x)$ the staggered fermion phase factors are denoted.
Further details on the definition of the action are given in \cite{Hel99}.
The tree level coefficients $c_1^F$ and $c_3^F$ appearing in $S_F$
have been fixed by demanding rotational invariance
of the free quark propagator at ${\cal O}(p^4)$ (``p4-action''). Moreover,
the 1-link term of this action has been modified by introducing ``fat''
links \cite{Blu97} with a weight $\omega = 0.2$. In the infinite
temperature limit the fat link term does not contribute to thermodynamic
observables nor to their cut-off dependence. Also at ${\cal O} (g^2)$
its effect has been found to be small \cite{Hel99}. It thus is expected
to be of little
importance for our current analysis, which is focused on the high
temperature behaviour of the pressure. The fat links are,
however, known to improve the flavour symmetry of the staggered fermion
action \cite{Blu97}.
Their contribution thus should become of significance in thermodynamic
calculations with light quarks close to $T_c$.

Even with our improved action cut-off effects  are still quadratic in the
lattice spacing {\it a}. In thermodynamic calculations this translates into
a quadratic dependence on the finite temporal extent of the lattice,
$(aT)^2 = 1/N_\tau^2$.
Compared to the standard gauge and staggered fermion actions these
contributions are, however, drastically reduced in magnitude.
For $\beta \rightarrow \infty$ the p4-action yields a much more
rapid approach to the continuum ideal gas limit
and deviates little from it already on lattices with rather small temporal
extent $N_\tau$. The deviations are smaller than 5\% already on
lattices with temporal extent $N_\tau = 6$.  Even with an improved
gauge sector this accuracy is reached with the standard staggered fermion
action only for $N_\tau \ge 16$. The cut-off distortion at some small
values of $N_\tau$ is given in Table~\ref{tab:free}.

\begin{table}[htb]
\begin{center}
\vspace{0.3cm}
\begin{tabular}{|r|c|c|c|c|c|}
\hline
$N_\tau$ & $p_G(N_\tau)/p_{\rm SB,G}$ &
\multicolumn{2}{|c|} {$p_F(N_\tau)/p_{\rm SB,F}$}&
\multicolumn{2}{|c|} {$(\epsilon_F(N_\tau)-3p_F(N_\tau))/T^4$} \\
\hline
~~&gauge action &p4-action&Naik-action &
p4-action&Naik-action \\
\hline
4~& 0.9284 & 0.5932 & 0.8070  & 1.3393 & 1.5991  \\
6~& 0.9925 & 0.9378 & 0.7718  & -0.0867 & 0.8079  \\
8~& 0.9983 & 0.9778 & 0.9303  & 0.0236 & 0.1156  \\
10~& 0.9994 & 0.9889 & 0.9809 & 0.0155 & 0.0202  \\
\hline
\end{tabular}
\end{center}
\caption{Cut-off dependence of the pressure on lattices
with temporal extent $N_\tau$. We separately give the gluonic contribution
and that for massless fermions normalized to the corresponding continuum
ideal gas values. The second column gives the result
for the tree level improved gauge action defined in Eq.~(2.2) and
the third column is for the p4-action defined in Eq.~(2.3). In the
third column we give for comparison the results for the ${\cal O} (a^2)$
improved Naik action \cite{Nai89}. The last two columns show the
violation of a basic thermodynamic identity for the ideal Fermi gas
(one flavour) due
to finite cut-off effects.}
\label{tab:free}
\end{table}

In the last two columns of Table~\ref{tab:free} we also show
$(\epsilon_F -3p_F)/T^4$ calculated by using the lattice versions of
standard thermodynamic relations, i.e.
$p/T = V^{-1}\ln Z$ and $ \epsilon = -V^{-1} \partial \ln Z / \partial (1/T)$.
The resulting integrals which have been evaluated numerically are
given for a free massless fermion gas,
\begin{eqnarray}
\lefteqn{ \frac{p_F(N_\tau)}{T^4}~=}\nn\\
&&\frac{3}{8} n_f N^4_\tau~\frac{1}{(2\pi)^3}~\int^{2\pi}_0 {\rm d}^3\vec{p}
\Bigg[N^{-1}_\tau
\sum^{N_\tau-1}_{n_0=0}~
\ln \left({\omega^2(\vec{p})+4f^2((2n_0+1)\pi/N_\tau)}\right)\nn\\
&& - \frac{1}{(2\pi)}~\int^{2\pi}_0~{\rm d}p_0~
\ln \left( {\omega^2(\vec{p})+4f^2(p_0)}\right)\Bigg]~~~,\\
\lefteqn{ \frac{\epsilon_F(N_\tau)}{T^4}~=}\nn\\
&&3 n_f N^4_\tau~\frac{1}{(2\pi)^3}~\int^{2\pi}_0 {\rm d}^3\vec{p}
\Bigg[N^{-1}_\tau
\sum^{N_\tau-1}_{n_0=0}~\frac{f^2((2n_0+1)\pi/N_\tau)}
{\omega^2(\vec{p})+4f^2((2n_0+1)\pi/N_\tau)}\nn\\
&& - \frac{1}{(2\pi)}\int^{2\pi}_0~{\rm d}p_0~\frac{f^2(p_0)}
 {\omega^2(\vec{p})+4f^2(p_0)}\Bigg]~~~,
 \end{eqnarray}
where the zero temperature contributions to $p_F/T^4$ and $\epsilon_F/T^4$
have been subtracted. Here the function
$\omega^2(\vec{p})\equiv 4\sum^3_{\mu=1}f^2(p_\mu)$ is introduced where 
in the case of the Naik and p4 action $f(p_\mu)$ is given by 
\begin{eqnarray}
f(p_\mu)&=&{9\over16}\sin(p_\mu)-{1\over48}\sin(3p_\mu)~~~{\rm(Naik-action)}\\
f(p_\mu)&=&~{3\over8}\sin(p_\mu)+{1\over48}~2~\sin(p_\mu)
\sum_{\nu\neq\mu}\cos(2p_\nu)~~~{\rm (p4-action)}.
\end{eqnarray}
In the temporal direction only
the discrete Matsubara modes $p_0=(2n_0+1)\pi/N_\tau$ 
with $n_0=0,~1,...,~(N_\tau-1)$ contribute.
The deviations from zero in $(\epsilon_F -3p_F)/T^4$ are a direct measure of the violation of
basic thermodynamic identities, valid for an ideal gas, due to finite cut-off
effects.
This also supports our preference for using the p4-action in
thermodynamic calculations rather than the Naik-action.

\section{The numerical calculation}

Our calculations for two and three flavour QCD have been performed
with quarks of mass $m_{u,d;L}=0.1$ on lattices of size $16^3\times 4$, i.e.
$m_{u,d}/T=0.4$. In addition we perform calculations with two light
quarks of mass $m_{u,d;L}=0.1$ and a heavier quark of mass $m_{s;L}=0.25$,
i.e. $m_s/T=1.0$.
To normalize the pressure and also in order to extract a physical
temperature scale additional {\it zero temperature} calculations
have been performed on symmetric $16^4$ lattices.
We have used the standard Hybrid R algorithm \cite{Got87} with a step size
$\Delta \tau \lsim \; m_{u,d;L}/2$ and a trajectory length $\tau=0.8$
 to update the gauge and fermion fields. On the $16^3\times 4$
lattice we collected
2000 to 3000 trajectories for values of the gauge coupling near the
critical  point and about 1000 trajectories for values away from it.
In the zero temperature simulations up to 800 trajectories were
generated to obtain a statistical error comparable to the finite
temperature calculations.

In a first step we determine the transition region for our choice
of parameters.
A pseudo-critical temperature has been determined on the $16^3\times 4$
lattice by locating the peak position of the susceptibility of the
Polyakov-loop and the chiral condensate, respectively. The location
of both peaks has been found to coincide within errors. The signal in
the chiral susceptibilities is, however, far more pronounced than in
the Polyakov loop susceptibility. The resulting
pseudo-critical couplings are given in Table~\ref{tab:tc}.

In order to determine a physical scale for our finite temperature
calculations we have calculated the heavy quark potential and also
performed spectrum calculations at $T=0$, i.e.
on the $16^4$ lattice. The heavy quark potential has been determined
in the usual way from smeared Wilson loops (for details see \cite{Bei99}).
From its long
distance behaviour we extract the string tension which then yields
$T_c/\sqrt{\sigma}$ and also defines a temperature scale
for our calculations of the pressure. The resulting
critical parameters are also given in Table~\ref{tab:tc}.

\begin{table}[htb]
\begin{center}
\vspace{0.3cm}
\begin{tabular}{|c|c|c|c|c|c|}
\hline
flavour content & $\beta_c$& $\sigma a^2$& $T_c/\sqrt{\sigma} $ &
$m_{\rm ps} a$ & $m_{\rm v} a$ \\ \hline
2 & 3.646~(4) & 0.271~(10) & 0.480~(10)& 0.958~(2) & 1.377~(25)\\
2+1 & 3.543~(2) & 0.271~(11) & 0.480~(10) & 0.962~(3) & 1.343~(20)\\
3 & 3.475~(2) & 0.283~(11) & 0.470~(9) & 0.967~(1) & 1.415~(15)\\
\hline
\end{tabular}
\end{center}
\caption{Pseudo-critical couplings, string tensions calculated at these
couplings and the resulting pseudo-critical temperatures for $n_f=2$ and
3 as well as QCD with two light and a heavier strange quark. In
addition the masses for the light pseudo-scalar and vector mesons are
given.}
\label{tab:tc}
\end{table}
We note that $T_c/\sqrt{\sigma}$ shows little dependence on the number of
flavours, although the present analysis certainly is not yet
indicative for the behaviour in the chiral limit where $T_c/\sqrt{\sigma}$
seems to be about 20\% smaller than our current result \cite{Kar99}.
Furthermore we have performed a spectrum calculation on a $16^4$
lattice at the pseudo-critical couplings. For the ratio of pseudo-scalar
and vector meson masses we find in all three cases,
${m_{\rm ps} / m_{\rm v}} \simeq 0.7$ (see Table~\ref{tab:tc}).
This also indicates that the quark masses used in our current analysis are
certainly too large to investigate in more detail the temperature
interval close to $T_c$. In the high temperature phase, however, the
dependence on the bare quark masses is strongly reduced \cite{Jos97,Ber97}
and a reliable estimate of the pressure in the chiral limit becomes 
possible.

\section{The pressure}

The free energy density, $f=-TV^{-1}\ln Z$, which in the thermodynamic
limit directly yields the pressure, $p=-f$, can be calculated using
the integral method \cite{Eng90}. As the logarithm of the partition
function is not directly accessible in a Monte Carlo calculation one
first differentiates the partition function, Eq.~(2.1),
with respect to the coupling $\beta\equiv 6/g^2$ and subsequently integrates
the resulting expectation values of the gauge action,
\beqn
{f\over T^4}\bigg|^\beta_{\beta_{0}}=-\biggl( {N_\tau \over N_\sigma}\biggr)^3
\int^\beta_{\beta_{0}}
d\beta^\prime \left(\langle S_G\rangle_0-\langle
S_G\rangle_T\right) \quad.
\label{fed}
\eqn
Here $N_\sigma$ and $N_\tau$ denote the spatial and temporal
extent of the finite temperature lattice, respectively. In the
limit $N_\sigma \rightarrow \infty$, Eq.~(\ref{fed}) gives the pressure,
$p(T)= -f(T)$. Experience from earlier calculations shows that aside from a
small temperature interval around $T_c$ this
limit is well approximated by $N_\sigma \ge 4N_\tau$.
Moreover, $\langle S_G\rangle_T$ is the expectation value of the 
gluonic part of the action at finite temperature.
The zero temperature contribution $\langle S_G\rangle_0$ calculated
on the $16^4$ lattice is subtracted to normalize the pressure
to zero at $T=0$. Strictly speaking Eq.~(\ref{fed}) gives the
difference between the ratios $f(T)/T^4$ calculated at two different
temperatures, $T=T(\beta)$ and $T_0=T(\beta_0)$. We choose the latter such
that $f(T_0)/T_0^4 \simeq 0$. Within our numerical accuracy this is the case
for $T_0 \simeq 0.6 T_c$.
Calculations of the differences of the action expectation values
have then been performed at about 15 different values of the coupling which
cover the temperature range $0.6\; T_c \le T \le 4\; T_c$.

\begin{figure}[htb]
\begin{center}
\epsfig{file=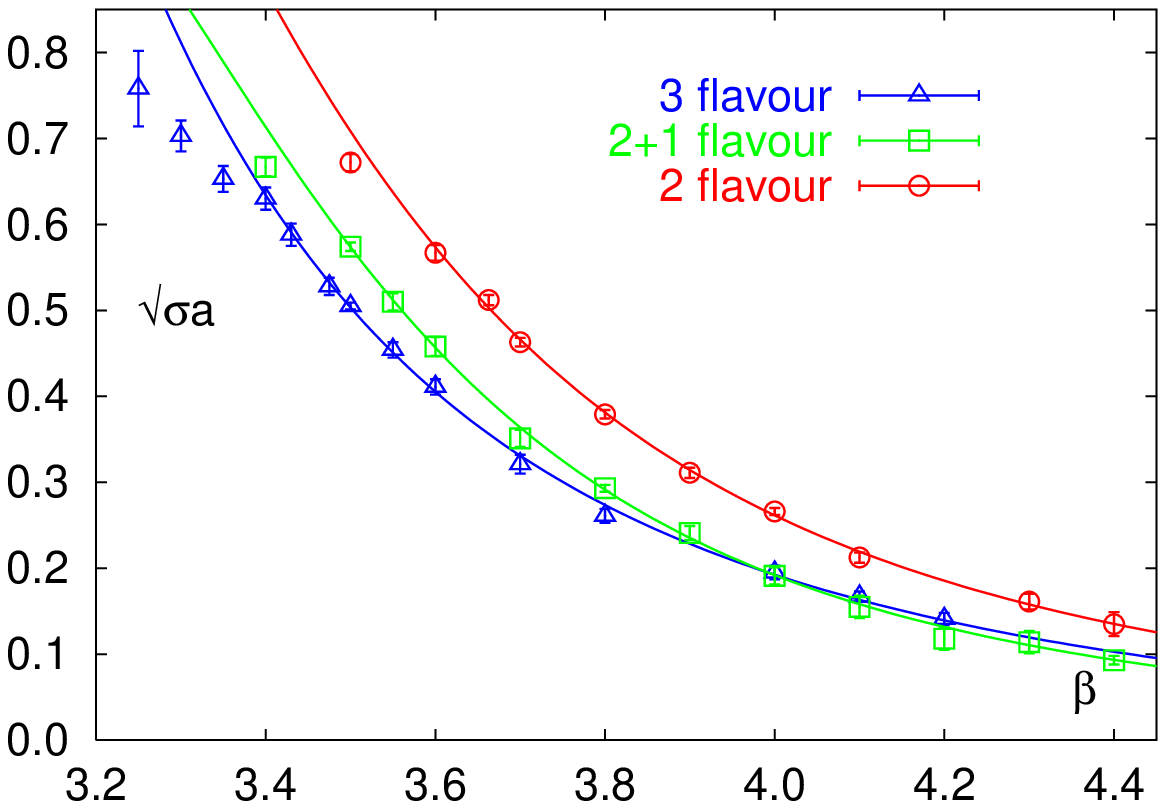,width=74mm}
\epsfig{file=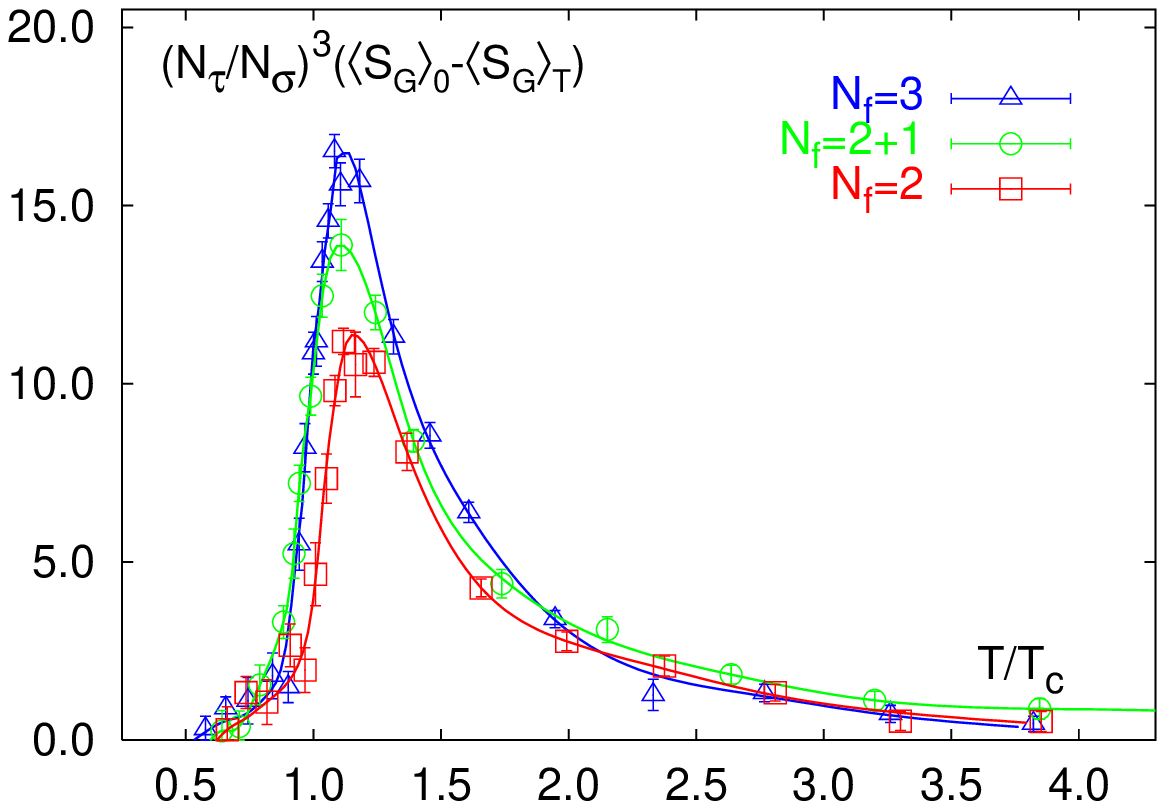,width=74mm}
\end{center}
\caption{String tension calculated on $16^4$ lattices (a) and
action differences from $16^4$ and $16^3\times 4$ lattices (b) for
$n_f=2$, 3 and (2+1).}
\label{fig_basic}
\end{figure}

The basic input for our analysis of the QCD thermodynamics thus consists
of a zero temperature calculation of the string tension, which is used
to fix the temperature scale, and a calculation of action differences.
The latter can then be integrated analytically
to give the pressure according to Eq.~(\ref{fed}).
These data are shown in Figure~\ref{fig_basic}.

For the interpolation of the string tension data
we use a renormalization group inspired ansatz \cite{All96},
\begin{equation}
\sqrt{\sigma a^2} (\beta) = R(\beta) (1+c_2 \hat{a}^2 (\beta)  +
c_4 \hat{a}^4 (\beta) )/c_0
\label{fit}
\end{equation}
with $\hat{a} \equiv R(\beta) / R(\bar{\beta})$.
The interpolation parameters are given in Table~\ref{tab:fit}.
Although this ansatz is, in principle, suitable for an extrapolation to the 
continuum limit we stress that
it is used here only as an interpolation for the string tension data
and as such is valid only in the interval indicated in Table~\ref{tab:fit}.

\begin{table}[htb]
\begin{center}
\vspace{0.3cm}
\begin{tabular}{|c|c|c|c|c|c|}
\hline
flavour content & $[\beta_{min},\beta_{max}]$& $\bar{\beta}$& $c_0$ &
$c_2$ & $c_4$ \\ \hline
2 & [3.6,4.4] & 3.70& 0.0570~(35)& 0.669~(208) & -0.0822~(1088)\\
2+1 & [3.5,4.4] & 3.60 & 0.0526~(32)& 1.026~(224) & -0.1964~(1065)\\
3 & [3.4,4.2] & 3.50 & 0.0448~(15)& 0.507~(115) & -0.0071~(677)\\
\hline
\end{tabular}
\end{center}
\caption{Fit parameters used for the interpolation of string tension
data.}
\label{tab:fit}
\end{table}

\begin{figure}
\begin{center}
\epsfig{ file=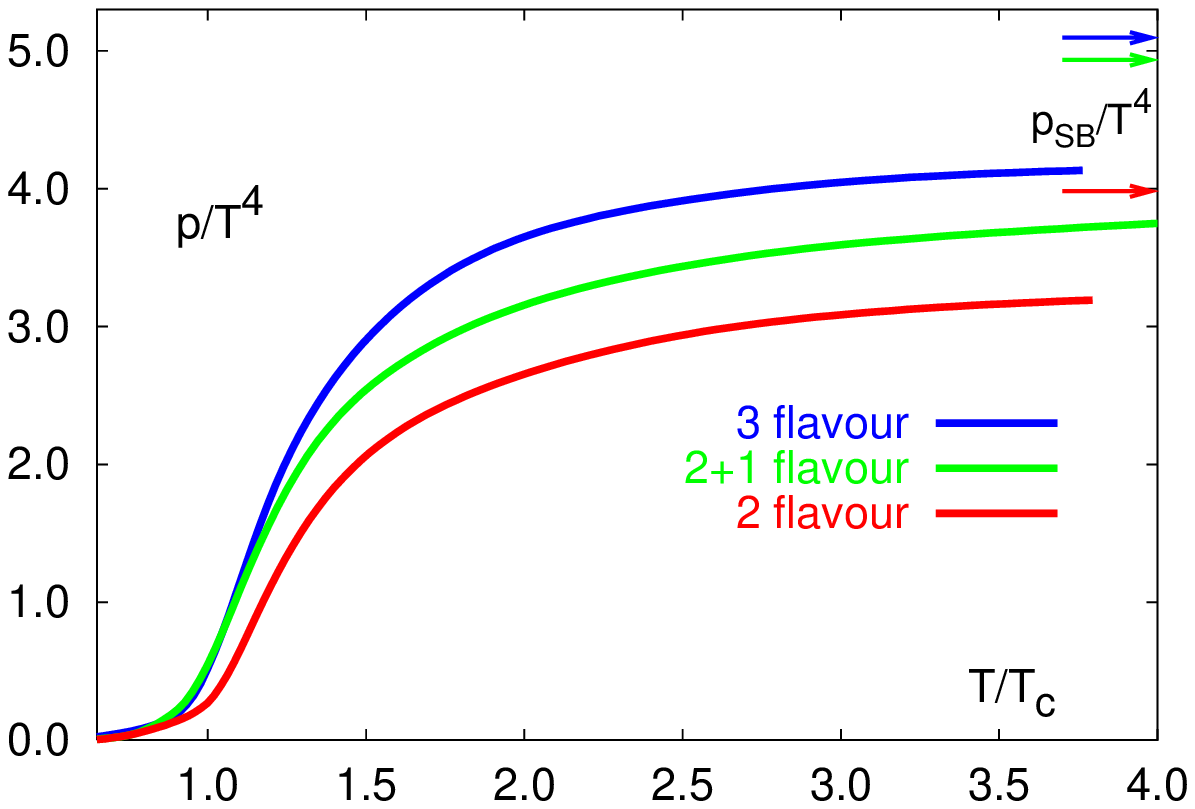,width=74mm}
\epsfig{file=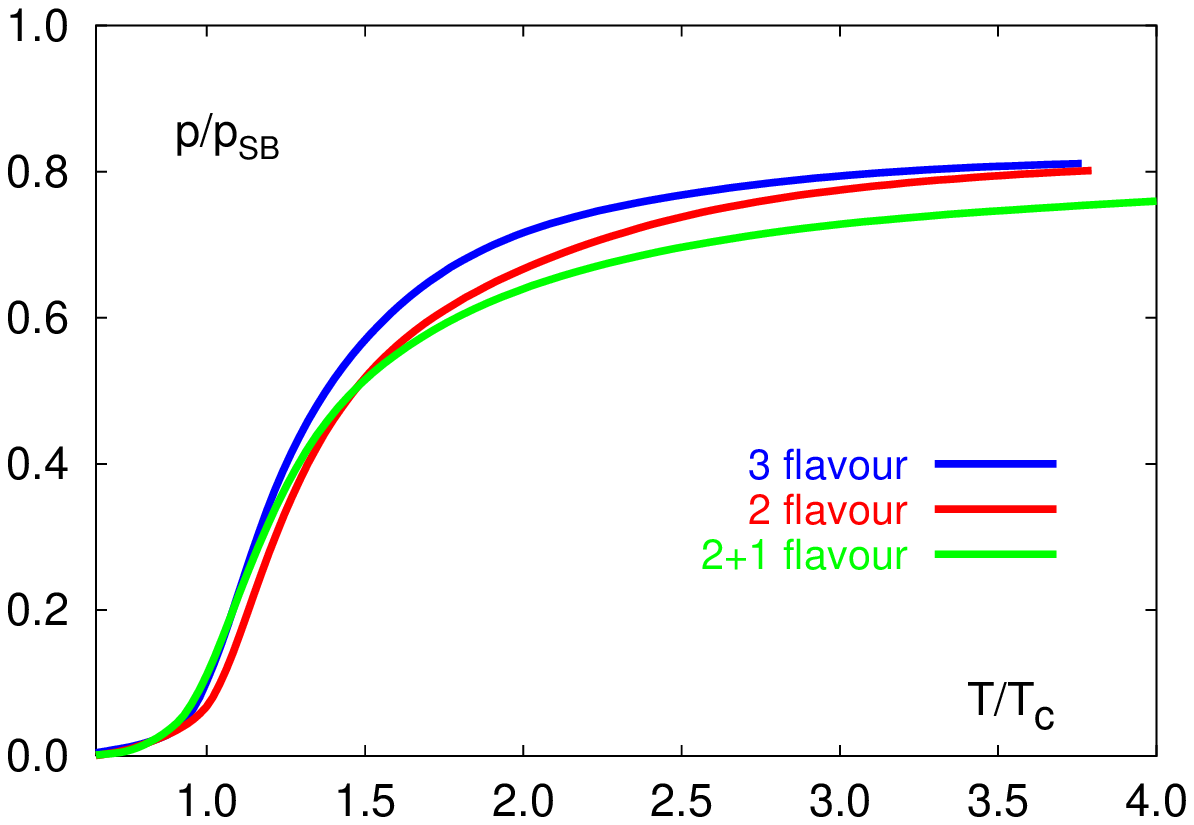,width=74mm}
\end{center}
\caption{The pressure for $n_f=2$, 2+1 and 3
calculated with the p4-action (a) and the normalized values $p/p_{\rm SB}$
(b).The arrows indicate the  continuum ideal gas limits for two and three
flavour QCD with quarks of mass $m/T=0.4$ as well as the case of two flavour
QCD with $m/T=0.4$ and an additional heavier quark of mass $m_s/T=1$.}
\label{fig:pressure}
\end{figure}

The systematic increase in the action differences with
increasing number of flavours visible in Figure~\ref{fig_basic} leads to
the increase of the pressure with increasing number of the
degrees of freedom,  which is apparent from Figure~\ref{fig:pressure}a,
where we show the pressure for $n_f=2$ and 3 as well as the (2+1)-flavour
case.  In fact, in the case of the simulations with two and three
light quarks, respectively, we observe that
this flavour dependence can almost completely be attributed to that of an
ideal quark-gluon gas,
\begin{equation}
{p_{\rm SB}\over T^4} = \biggl(16 + {21\over 2} g_f \biggr)
{\pi^2 \over 90} \quad .
\label{ideal}
\end{equation}
Here $g_f$ counts the effective number of degrees of freedom of a
massive Fermi gas. For a massless gas we have, of course, $g_f=n_f$.
In general we define
\beqn
g_f =\sum_{f=u,d,..} g(m_f/T)\quad ,
\eqn
with
\beqn
g(m/T)= {360\over 7\pi^4} \int_{m/T}^{\infty}{\rm d}x x
\sqrt{x^2-(m/T)^2} \ln\bigl(1+{\rm e}^{-x}\bigr)\quad .
\eqn
For the quark mass values used here one gets $g(0.4)=0.9672$ and $g(1)=0.8275$,
respectively. The correspondingly normalized curves are given in
Figure~\ref{fig:pressure}b. This indicates that in the presence of a
heavier quark the deviations of the pressure from the ideal gas value
is larger than in the massless limit. This is in qualitative agreement
with the observations made in \cite{Kog88}.

\vspace{0.5cm}

\section{Continuum Limit: An Estimate}

Our current analysis is restricted to a single temporal lattice size,
i.e. $N_\tau=4$. We are thus not yet in the position of performing
a complete extrapolation to the continuum limit. With our improved
action finite cut-off effects are, however, strongly reduced at high
temperature. We thus may attempt to give an estimate for the continuum
equation of state for massless QCD with two or three quark flavours
at temperatures not too close to $T_c$, e.g. $T \geq 2 T_c$.

The influence of a non-zero, yet small to moderately large quark
mass is small at high temperatures \cite{Jos97,Blu95,Ber97} and,
moreover, seems to be well described by that of an ideal Fermi gas,
Figure~\ref{fig:pressure}b. The ratio of Stefan-Boltzmann factors for
QCD with two and three light quarks of mass $m/T=0.4$ and massless
QCD is 0.981 and 0.978, respectively. Thus, this seems to be a minor
source for systematic deviations from the massless continuum limit.
The main source for systematic errors clearly still are finite
cut-off effects.

\begin{figure}
\begin{center}
\epsfig{file=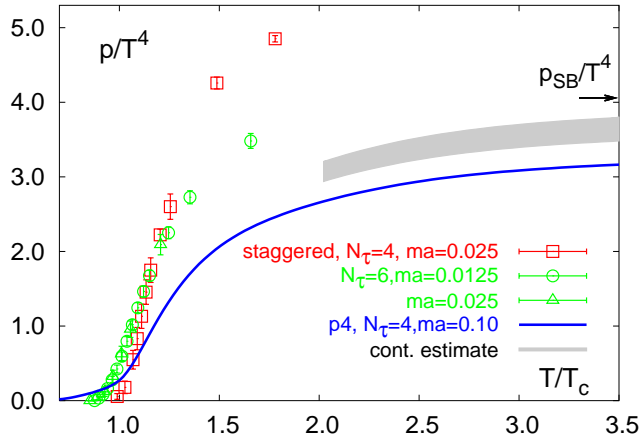,width=90mm}
\end{center}
\caption{The pressure for $n_f=2$.
Shown are results obtained with the p4-action on lattices with
temporal extent $N_\tau=4$ (line) as well as with the standard staggered
fermion action on $N_\tau=4$ (squares) and 6 (circles, triangles) lattices.
Also shown is an estimate of the continuum equation of state for
massless QCD (dashed band), based on the assumption that the systematic
error of the current analysis is $(15\pm 5)\%$.}
\label{fig:pres_cont}
\end{figure}

The analyzes of finite cut-off effects in the pure gauge theory have shown
that at temperatures $T\simeq~(2-4)~T_c$ the ideal gas calculations
correctly describe qualitative features of the cut-off
dependent terms. However, they overestimate their influence by roughly a
factor 2.
If this carries over to calculations with light quarks, which similar
to thermal gluons also acquire
a thermal mass of ${\cal O}(g(T)T)$ we may expect that the finite cut-off
distortion in our numerical calculations is also reduced by a similar factor.
From Table~\ref{tab:free} we find that in the ideal gas limit our improved
action leads to results which are 26\% and 29\% below the continuum value
for $n_f=2$ and 3, respectively.
Combined with the small systematic errors resulting from the use of
non-zero quark masses 
we thus expect that the continuum equation of state for massless QCD at
temperatures $T~\gsim~2T_c$ is about 15\% above the values
currently obtained in our analysis. This estimate for the continuum
limit is shown for two-flavour QCD in Figure~\ref{fig:pres_cont} where
we also show results from a calculation with the standard Wilson gauge
and staggered fermion action on lattices with temporal extent
$N_\tau=4$ and 6 \cite{Ber97}. These latter data lie substantially higher
which is in accordance with the larger cut-off effects for the unimproved
actions.

The lattice studies of the QCD free energy density, or equivalently the
pressure, indicate that deviations from the infinite temperature, ideal
gas limit are about (15-20)\% in the temperature interval
$2T_c~ <~ T~ <~ 4T_c$. This is quite similar to what has been found in
the pure gauge sector. We note that also the HTL resummed perturbative
calculations \cite{And99b,Bla99b} suggest similar deviations from ideal
gas behaviour. This does form a basis for more phenomenological
quasi-particle models for the QCD equation of state \cite{Pes96,Lev98}.

\section{Conclusions}

We have calculated the pressure in the high temperature phase of QCD
using improved gauge and fermion actions. Our analysis focuses on the
flavour dependence of the pressure and its dependence on a heavier (strange)
quark mass in the high temperature plasma phase. We find that the quark mass
dependence closely follows the pattern expected from the analysis of an ideal
Fermi gas. We observe, however, a significant reduction of the contribution
of strange quarks relative to that in an ideal gas. Interactions, which
at temperatures a few times $T_c$ lead to a strong reduction of the pressure
in QCD relative to that of an ideal gas thus also show a significant quark
mass dependence. We note, however, that the current analysis has been performed
with a fixed ratio $m_s/T$ rather than a fixed heavy quark mass. The
latter will be needed to arrive at quantitative predictions for the
plasma phase of QCD.

So far we only can give an estimate for the continuum extrapolated
pressure in the high temperature phase of QCD. To complete this analysis
and in particular in order to analyze the transition region and the
nature of the transition in (2+1) flavour QCD further calculations on
lattices with larger temporal extent and smaller quark masses are needed.
As the p4-action strongly reduces the cut-off dependence it provides a 
suitable framework for such more detailed studies with staggered fermions. 
Eventually one would
like to confirm the studies of the equation of state also within
another discretization scheme, in general with Wilson fermions.
Some work in this direction is underway \cite{Eji99}. However, also
here one will have to look for suitable improvements of the action
in order to reduce the cut-off dependence of thermodynamic observables.

\vspace{0.5cm}
\noindent
{\bf Acknowledgements:}

\medskip
\noindent
The work has been supported by the TMR network
ERBFMRX-CT-970122 and by the DFG under grant Ka 1198/4-1. FK thanks the
CERN Theory Devision for its hospitality. AP gratefully
acknowledges support through the research program "First Principle Calculations for
Hot Hadronic Systems" , Grant-in-Aide for Scientific Research , the Ministry of
Education, Science and Culture, Japan, No 11694085 and thanks for the
hospitality at Hiroshima
University where this work has been finalized.


\end{document}